\documentclass[journal=jpclcd]{achemso}
\usepackage{amsfonts,amssymb,amsmath,latexsym}
\usepackage{graphicx}
\usepackage{xcolor}
\usepackage{setspace}
\usepackage[utf8]{inputenc}
\usepackage{graphicx}
\usepackage{multirow}
\usepackage{tabularx}
\usepackage{soul}
\newcommand{\lcpq}{Laboratoire de Chimie et Physique Quantiques, CNRS, Universit\'e de Toulouse, UPS, 118 route de Narbonne, F-31062 Toulouse, France}
\newcommand{\warwick}{Department of Physics, University of Warwick, Coventry, CV4 7AL, United Kingdom}
\newcommand{\sanseb}{Polimero eta Material Aurreratuak: Fisika, Kimika eta Teknologia saila, Kimika Fakultatea, Euskal Herriko Unibertsitatea, UPV/EHU, and Donostia International Physics Center (DIPC), Donostia, P.K. 1072, 20080, Spain}
\newcommand{\valencia}{Departamento de Química Física, Universitat de València, Dr. Moliner 50, Burjassot, 46100, Spain}
\newcommand{\etsf}{European Theoretical Spectroscopy Facility (ETSF)}

\title[The Emergence of the Hexagonal Lattice in Two-Dimensional Wigner Fragments]{The Emergence of the Hexagonal Lattice in Two-Dimensional Wigner Fragments}
\label{title}

\author{Miguel Escobar Azor}
\affiliation{\warwick}
\affiliation{\lcpq}
\affiliation{\etsf}
\author{Amer Alrakik}
\affiliation{\lcpq}
\affiliation{\etsf}
\author{Louan de Bentzmann}
\affiliation{\lcpq}
\author{Xabier Telleria-Allika}
\affiliation{\sanseb}
\author{Alfredo S\'anchez de Mer\'as}
\affiliation{\valencia}
\author{Stefano Evangelisti}
\affiliation{\lcpq}
\author{J.~Arjan Berger}
\email{arjan.berger@irsamc.ups-tlse.fr}
\affiliation{\lcpq}
\affiliation{\etsf}

\begin{document}

\begin{tocentry}
\includegraphics{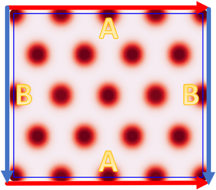}
\end{tocentry}

\begin{abstract}
\label{abstract}
At very low density, the electrons in a uniform electron gas spontaneously break symmetry and form a crystalline lattice called a Wigner crystal.
But which type of crystal will the electrons form?
We report a numerical study of the density profiles of fragments of Wigner crystals from first principles. 
To simulate Wigner fragments we use Clifford periodic boundary conditions and a renormalized distance in the Coulomb potential. 
Moreover, we show that high-spin restricted open-shell Hartree-Fock theory becomes exact in the low-density limit.
We are thus able to accurately capture the localisation in two-dimensional Wigner fragments with many electrons. 
No assumptions about the positions where the electrons will localise are made. The density profiles we obtain emerge naturally when we minimise the total energy of the system.  
We clearly observe the emergence of the hexagonal crystal structure which has been predicted to be ground-state structure of the two-dimensional Wigner crystal.
\end{abstract}

\maketitle

\renewcommand{\baselinestretch}{1.5}

In 1934 Eugene Wigner argued that in a uniform gas of electrons the Coulomb repulsion is dominant with respect to the kinetic energy when the average density is sufficiently low~\cite{Wigner_1934}.
Therefore, he predicted that the system would naturally break its translational invariance with the electrons localizing at fixed positions in space thus forming a crystal.
\emph{But which type of crystal would the electrons form?}
It was later predicted that in one dimension the electrons would form a linear chain, in two dimensions they would form a hexagonal (or triangular) lattice and in three dimensions a body-centred cubic lattice.
All these predictions
%, both using classical point charges and electrons, 
were based on the comparison of the energies for a few known crystal structures~\cite{Wigner_1934,Fuchs_1935,Sholl_1967,Bonsall_1977,Tanatar_1989,Rapisarda_1996,Hasse_1991,Drummond_2004,Drummond_2009,Alves_2021}.
For example, in two dimensions it can be shown that the hexagonal lattice has a lower energy than the square lattice.
However, this leaves open the possibility that there might be another, less trivial, lattice structure that has a lower energy than the hexagonal lattice~\cite{GiulianiVignale}.
In this work we make an important step towards the prediction of the lattice structure of Wigner crystals \emph{without} making any assumptions about the lattice structure.
We achieve this by spatially confining the electrons at very low (average) density and minimizing the total energy using a variational approach to find the electronic wave function of the ground state.
The electronic density corresponding to this wave function then corresponds to the lowest-energy configuration of the electrons.
In this work we will study fragments of Wigner crystals in which the translational symmetry of the crystal is retained.
We will therefore refer to these systems as Wigner fragments~\cite{Alrakik_2023}.
%Since the number of electrons that are spatially confined is too small to speak of Wigner crystals we will refer to them as Wigner fragments~\cite{Alrakik_2023}.
We will show that, in two dimensions, when increasing the number of electrons, there is a clear emergence of the hexagonal crystal structure in the Wigner fragments.

Recently, a two-dimensional Wigner crystal has been observed experimentally by Smolenski and coworkers~\cite{Smolenski_2021}.
However, they did not determine the lattice structure.
Two-dimensional crystal lattices consisting of electrons have also be obtained in other experiments, mainly by using magnetic fields or Moir\'e superlattices~\cite{Lozovik_1975,Grimes_1979,Andrei_1988,Goldman_1990,Williams_1991,Ye_2002, Chen_2006, Jang_2017,Regan_2020,Tang_2020,Xu_2020,Li_2021}. 
In the literature, these systems are also frequently called Wigner crystals.
%The properties of Wigner crystals have been studied extensively in condensed matter physics and have important implications for the understanding of the behavior of electrons in low-dimensional systems.
Closely related to the Wigner fragments studied here, are Wigner molecules, in which the electrons are typically confined by an external potential which breaks the translational invariance of the electron gas
~\cite{Cioslowski_2006,Ellenberger_2006,Yannouleas_2007,Mendl_2014,Cioslowski_2017,Cioslowski_2017_JCP,Egger,Diaz-Marquez_2018,Telleria_2022,Escobar_2019}.
Wigner molecules have also been observed experimentally~\cite{Pecker2013,Mendez-Camacho2022_1,Mendez-Camacho2022_2,Thakur_2022}.

In some recent works, we presented Clifford periodic boundary conditions and a renormalized distance as a simple and efficient approach to describe periodic Coulomb systems~\cite{Tavernier_2020_JPCL,Tavernier_2021,Alves_2021,Escobar_JCP_2021}.
Amongst other applications, we applied our approach to describe the Wigner localisation of two electrons in one, two, and three dimensions using a regular grid of gaussian basis functions in a large supercell that has the topology of a Clifford torus~\cite{Escobar_JCP_2021}.
In this work we present a general approach that can treat Wigner fragments with, in principle, any number of electrons.

\begin{figure}[!h]
    \centering
    \includegraphics[width=0.75\columnwidth]{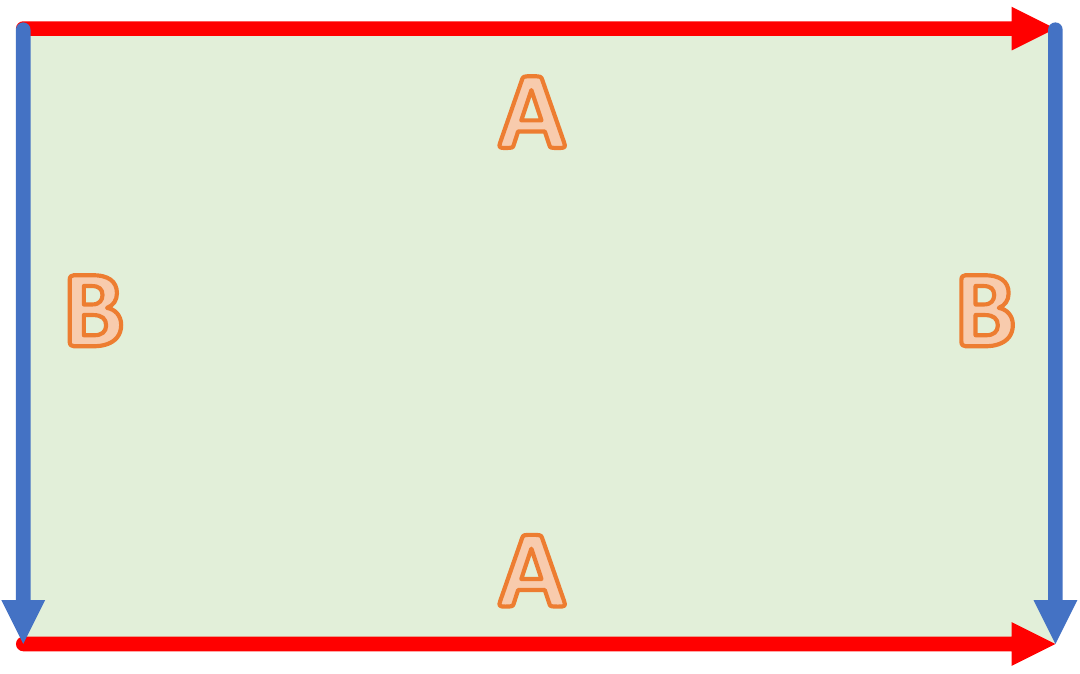}
    \caption{A representation of a 2-dimensional Clifford torus. Its fundamental polygon is a rectangle.
    In a Clifford torus opposite sides are joint without deformation.}
    \label{CSC}
\end{figure}

We consider an electron gas confined to a two-dimensional Clifford torus.
To be more precise, we define a rectangular 2-dimensional supercell containing a fragment of an electron gas, and then modify its topology into that of a torus by joining opposite sides of the supercell \emph{without deformation}. 
This procedure yields a supercell that has the topology of a 2-dimensional Clifford torus, which is a flat, closed 2-dimensional real Euclidean space embedded in a 2-dimensional complex Euclidean space.
Therefore, we refer to this supercell as the Clifford supercell.
The most important property of the Clifford torus is its flatness, i.e., it has zero gaussian curvature everywhere.
As an illustration, we report in Fig.~\ref{CSC} the rectangular representation of a 2-dimensional Clifford torus.
More details of our approach based on Clifford periodic boundary conditions can be found in Ref.~\cite{Escobar_JCP_2021}.

The Hamiltonian of $N$ electrons confined to the Clifford torus, in Hartree atomic units, is given by 
\begin{equation}
    \hat{H}=-\frac{1}{2}\sum_{i=1}^{N}\nabla^2_i + \frac12\sum_{i=1}^N \sum_{\substack{j=1\\j\neq i}}^N
    \frac{1}{r_{ij}},
    \label{Ham}
\end{equation}
where the first term on the right-hand side is the kinetic-energy operator of the electrons and the second term is the Coulomb repulsion between the electrons.
The Coulomb potential is defined in terms of the Euclidean distance $r_{ij}$ which is the distance between electrons $i$ and $j$ in the embedding space of the Clifford torus.
This ensures that all derivatives of the Coulomb potential are continuous everywhere in the Clifford supercell.
We note that this is not the case if $r_{ij}$ would be defined as the distance \emph{on} the Clifford torus~\cite{Tavernier_2021}.
In two dimensions this renormalized distance is defined as
\begin{equation} 
    \label{euclidian_distance}
    r_{ij} = \frac{1}{\pi}\sqrt{L_x^2 \sin^2  \left[\frac{\pi}{L_x} (x_i-x_j ) \right] + L_y^2 \sin^2  \left[\frac{\pi}{L_y} (y_i -y_j) \right]},
\end{equation}
where $L_x$ and $L_y$ are the lengths of the edges of the Clifford supercell.
\textcolor{red}{We note that in the limit of infinite system size, results obtained with the renormalised distance should be equal to those obtained with the usual distance~\cite{Valenca_2019,Evangelisti_2022}}

To obtain the lowest-energy configuration of the electrons at low electronic density, one could solve the time-independent Schr\"odinger equation with the Hamiltonian given in Eq.~\eqref{Ham}.
From the wave function corresponding to the lowest energy eigenvalue, the positions where the electrons localise can be obtained.
However, such a calculation is numerically feasible only for very few electrons.
Therefore, we have to find a numerically more efficient approach that can capture the electron localisation at low density.
We make use of two properties of the uniform electron gas in the low-density limit.
In this limit
1) the electrons behave as a single particle, since fixing the position of one electron uniquely determines the positions of all the others~\cite{Malet_2013,Mendl_2014};
2) the spin configuration of the wave function becomes irrelevant, i.e., all spin configurations become degenerate.
As a consequence, at very low density, we can focus on the high-spin state, which has the maximum value for the spin projection $|S_z|$.
The advantage of this spin state is that it can be described with a single Slater determinant by fixing the position of a single electron.
%Without loss of generality, we can now fix the position of one electron and only one Slater determinant
A Schr\"odinger equation for which the solution is a single Slater determinant can be solved with the Hartree-Fock method.
This means that the contributions from interactions beyond those included in the Hartree-Fock method become negligible in the low-density limit for the high-spin state.
Therefore, in the following we will use the restricted open-shell Hartree-Fock (ROHF) approach.

In practical calculations we have to project the Hartree-Fock equations onto a basis set. 
We use a gaussian basis that is distributed regularly in the Clifford supercell to perform our numerical calculations.
All gaussians have the same exponent that is adapted to the size of the supercell in such a way to have an overlap between
nearest neighbours that is sufficiently large to accurately describe the wave function but not too close to unity, in order to avoid numerical problems
due to quasi-linear dependence. 
We have shown that the overlap between nearest neighbours is proportional to the parameter $\xi = \alpha \delta^2$ in which $\alpha$ is the exponent of the gaussian and $\delta$ the nearest neighbour distance between the gaussians~\cite{Brooke_2018}.
Unless stated otherwise, we used 400 gaussians on a 20$\times$20 grid and $\xi= 0.8$.
We have implemented our approach in the DALTON software package~\cite{DALTON}.
All calculations on two-dimensional (2D) Wigner crystals have been performed at a Wigner-Seitz radius $r_s$ equal to $105$.
This is well above the theoretically predicted value of $r_s = 31$ for the Fermi liquid to Wigner crystal phase transition in two dimensions~\cite{Tanatar_1989,Rapisarda_1996,Drummond_2009}.
\textcolor{red}{We note that there are results obtained in one dimension that suggest that Wigner crystallisation could take place at smaller values of $r_s$.~\cite{Ostili_2021}}
Without loss of generality, we accelerated the convergence of the calculations by adopting the following protocol:
1) we performed a first calculation in which we fix one electron by adding a small positive charge in the origin of the Clifford supercell;
2) we then performed a second calculation without the small positive charge using the density of the first calculation as a starting point.

\begin{figure*}[t]
    \centering
    \begin{minipage}{0.25\textwidth}
    \centering
    \includegraphics[width=1.0\textwidth]{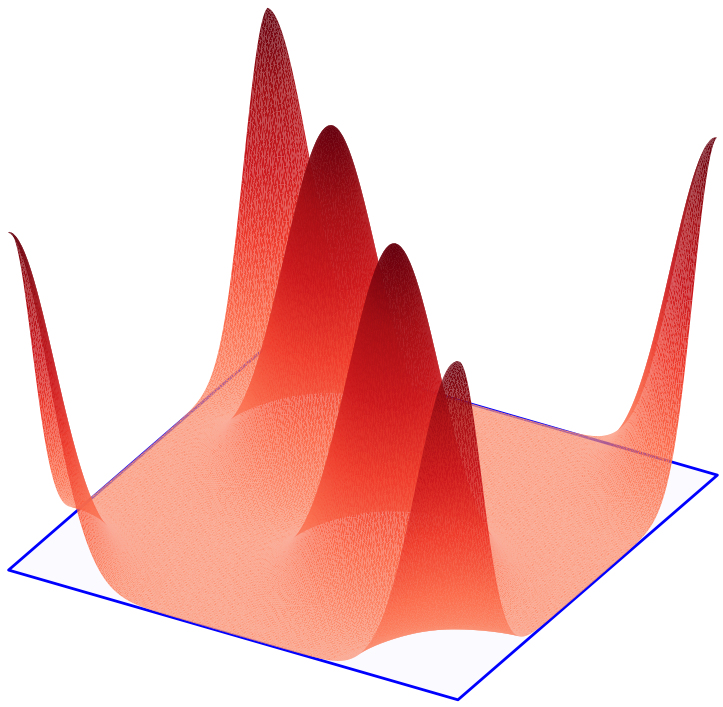}
    \end{minipage}
    \begin{minipage}{0.25\textwidth}
    \centering
    \includegraphics[width=1.0\textwidth]{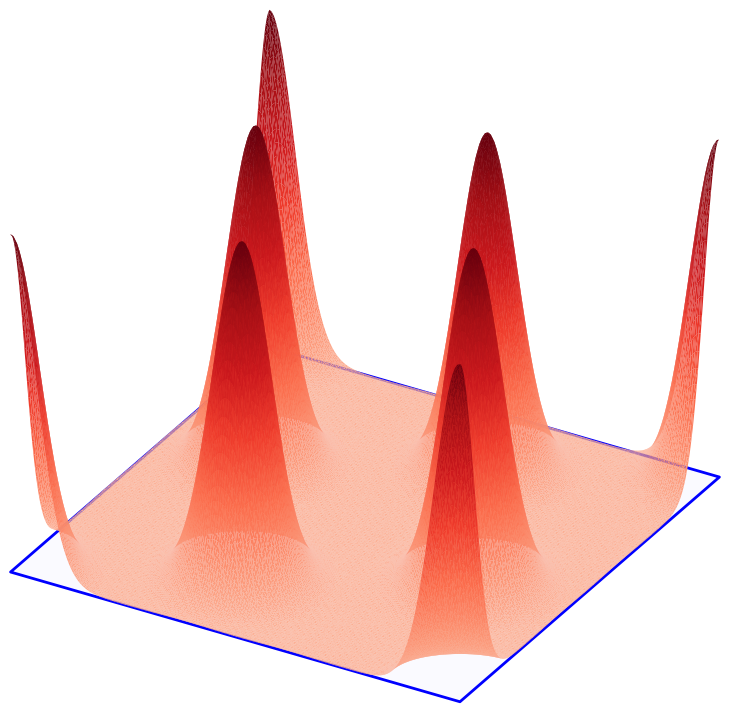}
    \end{minipage}
    \begin{minipage}{0.25\textwidth}
    \centering
    \includegraphics[width=1.0\textwidth]{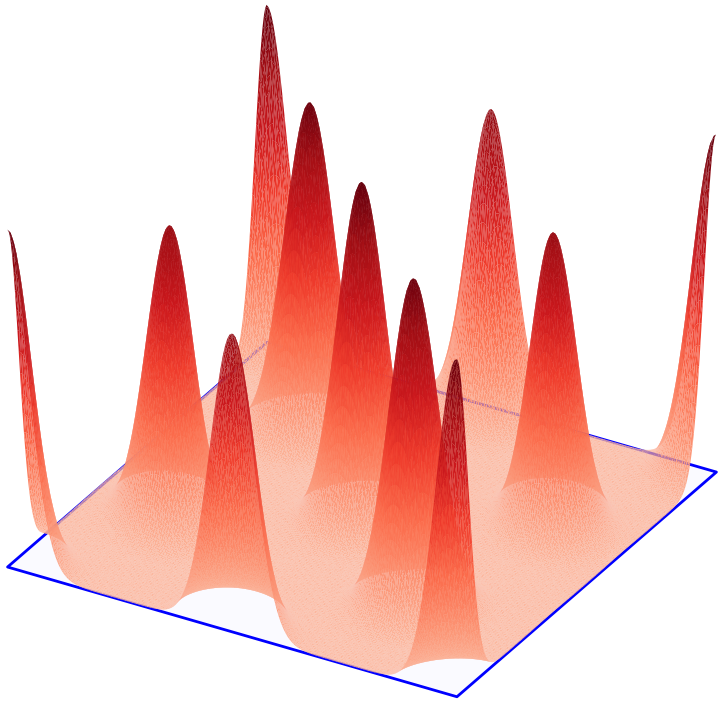}
    \end{minipage}
\\
    \begin{minipage}{0.25\textwidth}
    \centering
    \includegraphics[width=1.0\textwidth]{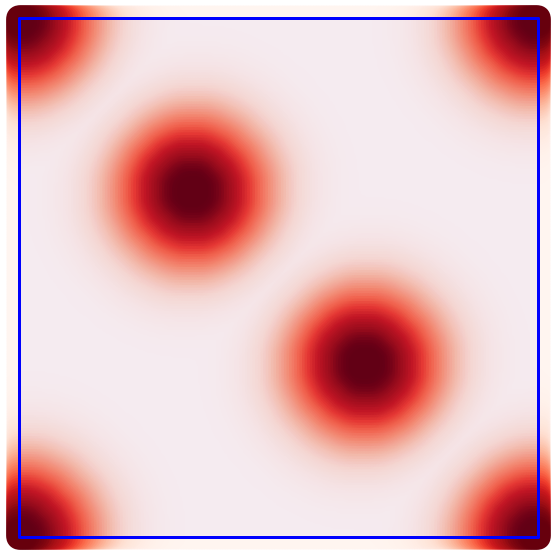}
    \end{minipage}
    \begin{minipage}{0.25\textwidth}
    \centering
    \includegraphics[width=1.0\textwidth]{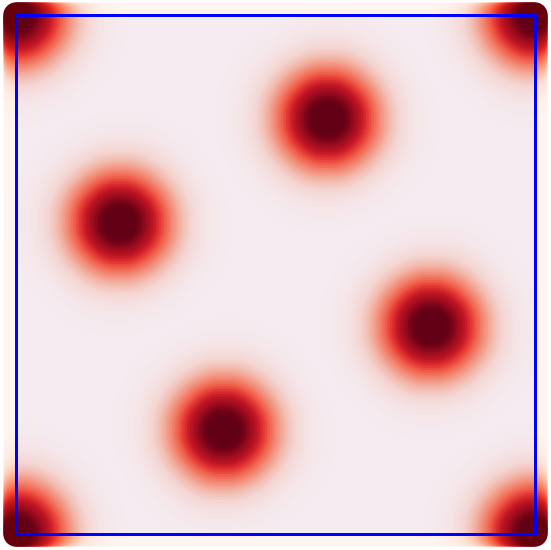}
    \end{minipage}
    \begin{minipage}{0.25\textwidth}
    \centering
    \includegraphics[width=1.0\textwidth]{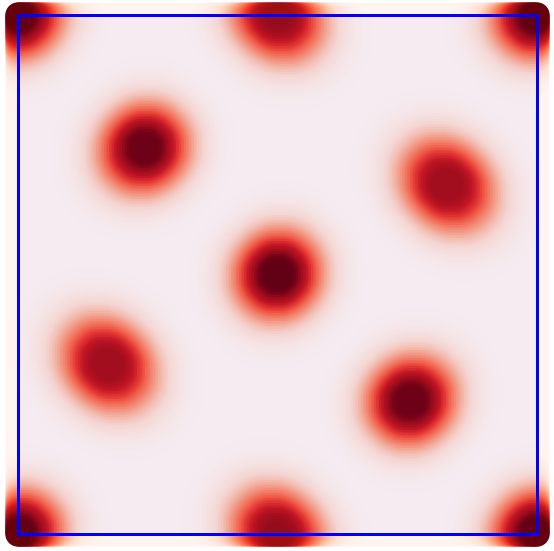}
    \end{minipage}
    \caption{The density profiles of 3 (left panel), 5 (middle panel) and 7 (right panel) electrons confined to a 2D square Clifford supercell at $r_s =105$.
    The profiles on the top show a view from the front while the profiles at the bottom show a view from above.}
    \label{357el}
\end{figure*}

Before reporting the results on 2D Wigner fragments, we note that we have done preliminary work on one-dimensional (1D) Wigner fragments.
In particular, we verified numerically that in the low-density limit the wave function indeed becomes mono-determinantal.
We achieved this by performing accurate multi-configurational calculations using the complete-active-space self-consistent field (CASSCF) method~\cite{Hegarty_1979} on 2 electrons confined to a Clifford supercell of length $L$.
We found that at extremely low density ($r_s= 25\,000$) the occupation numbers of the lowest two spin-orbitals are 0.99975 using a CAS(2,4).
This confirms that in the low-density limit the ground-state wave function can be described by a single Slater determinant.

In the left panel of Fig.~\ref{357el} we report the electron density of a Wigner fragment with three electrons confined to a square Clifford torus.
We observe that the electrons localise on the diagonal of the Clifford supercell.
The three-electron 2D system thus behaves as a 1D system with length $L=\sqrt{L_x^2+L_y^2}$.
We note that the reported solution is degenerate with another solution in which the electrons localise on the other diagonal of the Clifford supercell.
In the following, for the sake of simplicity, we will avoid discussing the degenerate solutions we have obtained, since the density profiles of all degenerate solutions are equivalent.
\begin{figure*}[t]
    \centering
    \begin{minipage}{0.25\textwidth}
    \centering
    \includegraphics[width=1.0\textwidth]{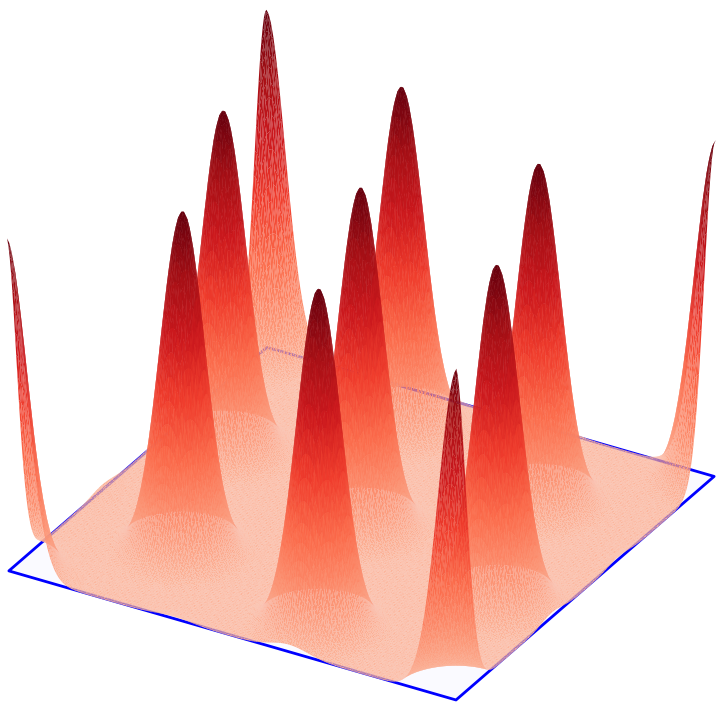}
    \end{minipage}
    \begin{minipage}{0.25\textwidth}
    \centering
    \includegraphics[width=1.0\textwidth]{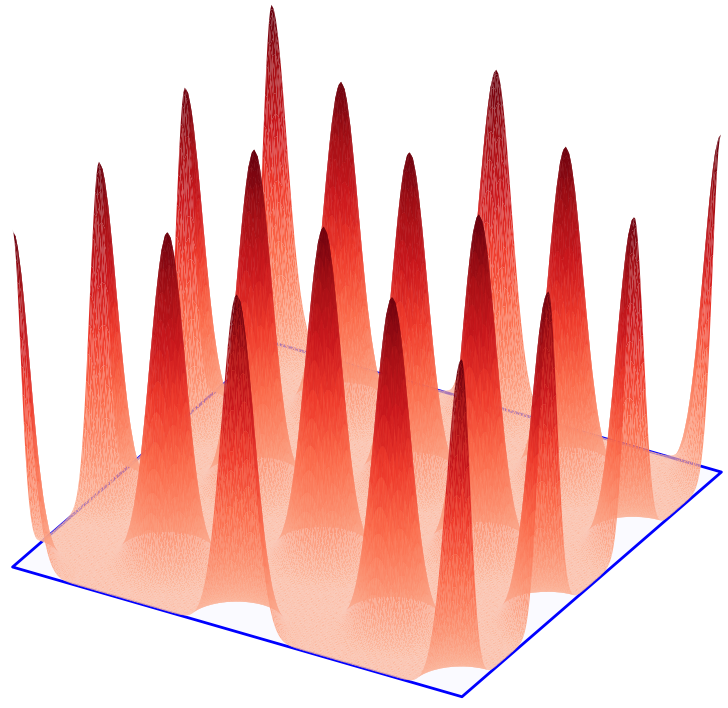}
    \end{minipage}
    \begin{minipage}{0.25\textwidth}
    \centering
    \includegraphics[width=1.0\textwidth]{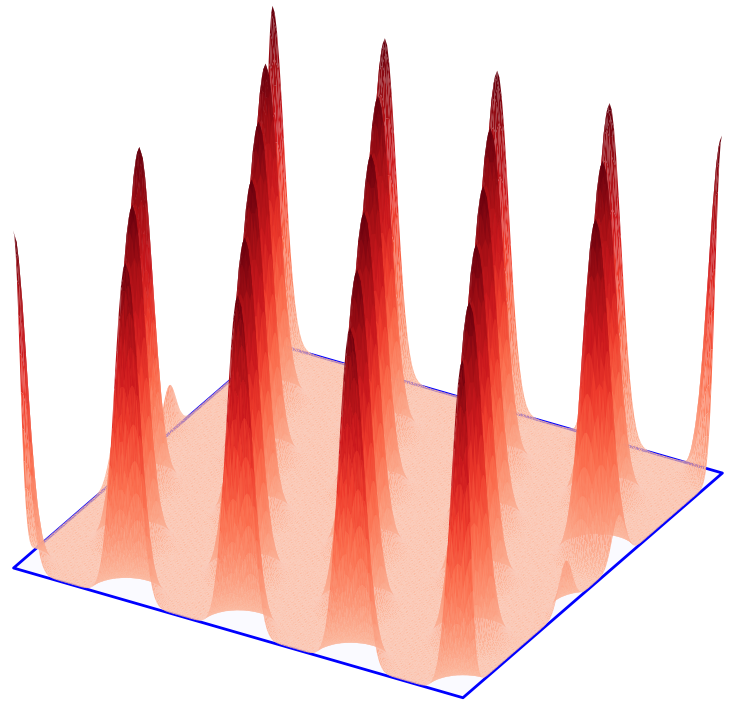}
    \end{minipage}
\\
    \begin{minipage}{0.25\textwidth}
    \centering
    \includegraphics[width=1.0\textwidth]{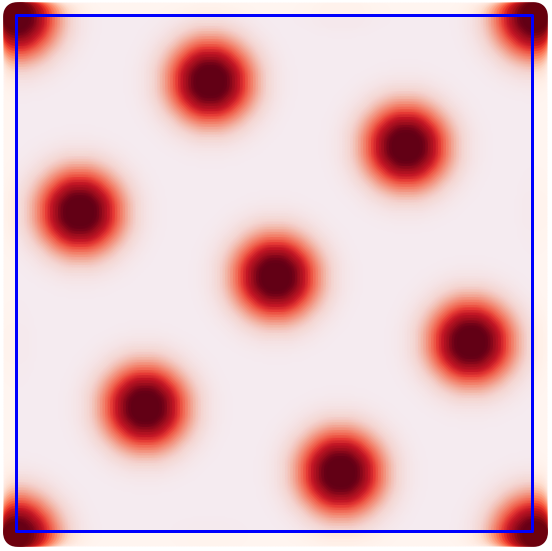}
    \end{minipage}
    \begin{minipage}{0.25\textwidth}
    \centering
    \includegraphics[width=1.0\textwidth]{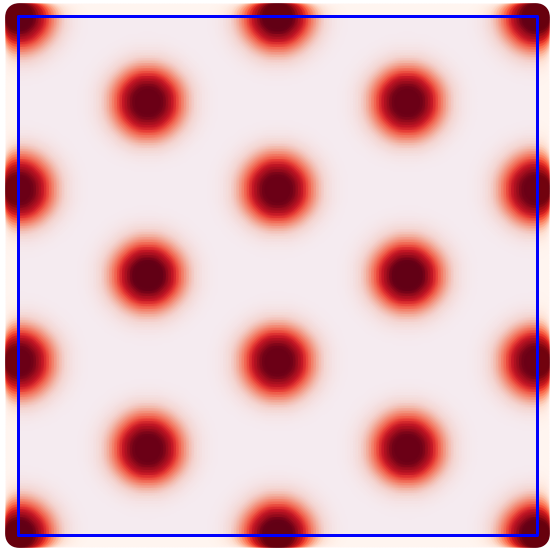}
    \end{minipage}
    \begin{minipage}{0.25\textwidth}
    \centering
    \includegraphics[width=1.0\textwidth]{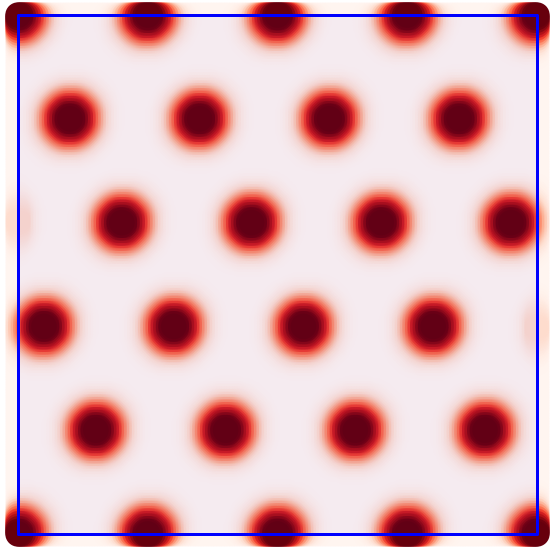}
    \end{minipage}
    \caption{The density profiles of 8 (left panel), 12 (middle panel) and 20 (right panel) electrons confined to a 2D square Clifford supercell at $r_s =105$.
    The profiles on the top show a view from the front while the profiles at the bottom show a view from above.}
    \label{81220el}
\end{figure*}

In the middle panel of Fig.~\ref{357el} we report the density for 5 electrons confined to a square Clifford supercell.
We observe that the electrons localise on two parallel lines, with each line joining a vertex to the center of an edge of the Clifford supercell.
For both 3 and 5 electrons all electrons are equivalent but this is not always the case.
For example, for 7 electrons confined to a square Clifford supercell there are 4 inequivalent electrons.
We report this result in the right panel of Fig.~\ref{357el}. 
There are 4 types of density distributions around the center of a localised electron that differ slightly in their amplitudes and widths.
Moreover, some density distributions are anisotropic.

An interesting localisation pattern emerges in a Wigner fragment with 8 electrons confined to a square Clifford supercell reported in the left panel of Fig.~\ref{81220el}.
We see that the localised electrons form a distorted hexagonal lattice.
Approximate hexagonal structures also emerge naturally in other square Clifford tori with more electrons such as $N=12$ and $N=20$.
We report those density profiles in middle and right panels of Fig.~\ref{81220el}.

\begin{figure}[b]
    \centering
    \begin{minipage}{0.225\textwidth}
    \centering
    \includegraphics[width=1.0\textwidth]{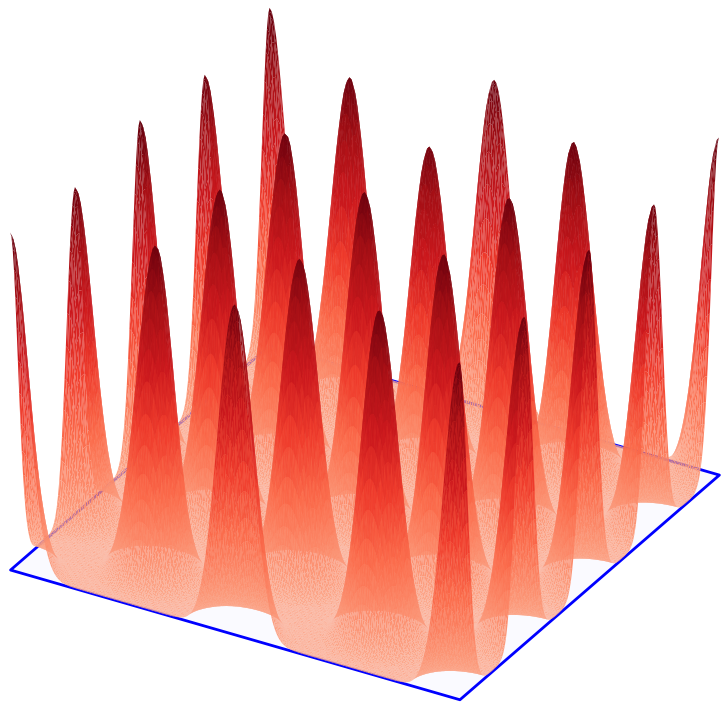}
    \end{minipage}
    \begin{minipage}{0.225\textwidth}
    \centering
    \includegraphics[width=1.0\textwidth]{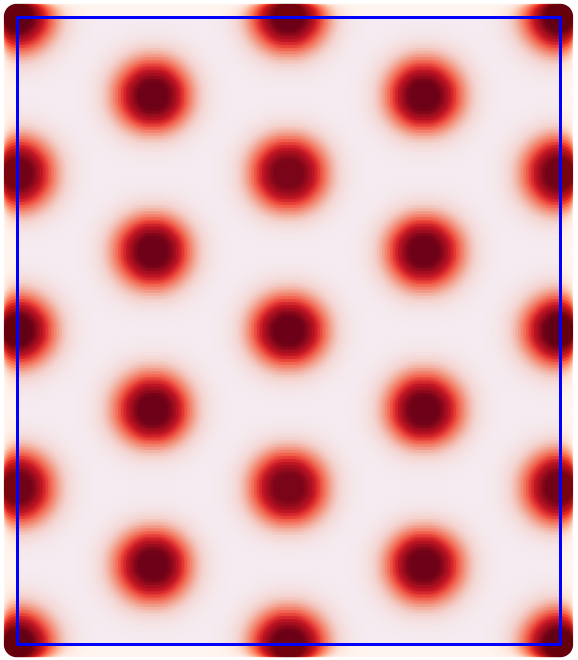}
    \end{minipage}
    \caption{The density profiles of 16 electrons confined to a 2D square Clifford supercell commensurable with the hexagonal lattice at $r_s =105$.
    The profile on the left shows a view from the front while the profile on the right shows a view from above.}
    \label{16el_hex_nd}
\end{figure}

\begin{figure}[t]
    \centering
    \begin{minipage}{0.225\textwidth}
    \centering
    \includegraphics[width=1.0\textwidth]{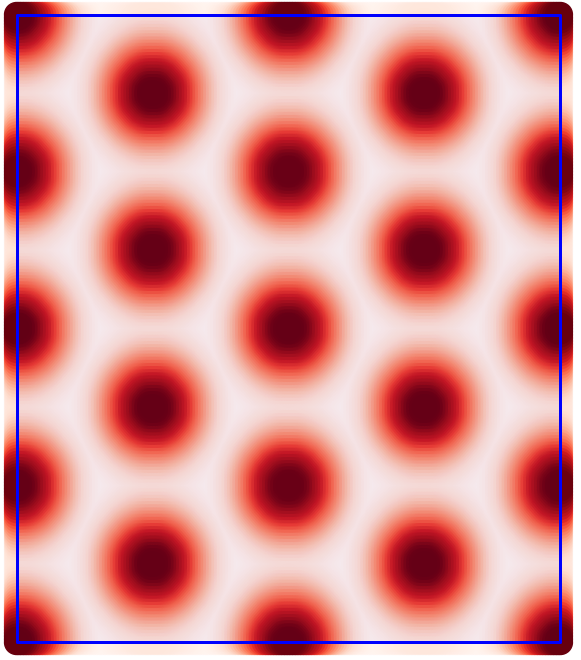}
    \end{minipage}
    \begin{minipage}{0.225\textwidth}
    \centering
    \includegraphics[width=1.0\textwidth]{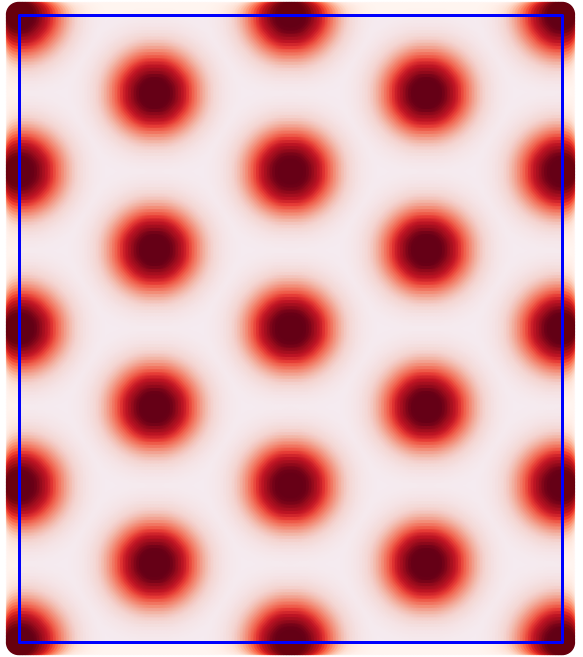}
    \end{minipage}
    \begin{minipage}{0.225\textwidth}
    \centering
    \includegraphics[width=1.0\textwidth]{2D_16_hex_topview.png}
    \end{minipage}
    \caption{The density profiles (view from above) of 16 electrons confined to a 2D square Clifford supercell commensurable with the hexagonal lattice for various values of $r_s$. 
    Left panel: $r_s = 25$; middle panel: $r_s$ = 50, right panel: $r_s$ = 105}
    \label{16el_hex_rs}
\end{figure}

In the above calculations the edges of the Clifford supercell were fixed to be equal, i.e., $L_x = L_y$.
For such a Clifford supercell it is not possible to find a perfect hexagonal lattice because it is incommensurable with a square for any number of electrons.
To verify if indeed a perfect hexagonal lattice will emerge, we now choose the ratio $L_y / L_x$ and the number of electrons to be commensurable with the hexagonal lattice.
To achieve this, we performed a calculation of 16 electrons confined to a rectangular Clifford supercell with $L_y / L_x = \sqrt{3}/2$.
We used 418 gaussians on a 22$\times 19$ grid.
The corresponding density profile can be found in Fig.~\ref{16el_hex_nd}.
We obtain a perfect hexagonal lattice that demonstrates the emergence of the hexagonal structure in a uniform electron gas at low density.
\textcolor{red}{It is also interesting to study the density profile of the Wigner phase as a function of $r_s$.
Therefore, in Fig.~\eqref{16el_hex_rs} we report the density profiles corresponding to $r_s = 25$ and $r_s = 50$ together with the profile for $r_s = 105$.
We observe that the relative positions of the electrons in the Wigner phase are independent of $r_s$ . 
However, at $r_s = 25$ we clearly see that the peaks are broader than those corresponding to $r_s = 105$}

We have calculated the density profiles of Wigner fragments for $N$ electrons in a square Clifford supercell for all values of $N>1$ up to $N=20$ (for $r_s = 105$).
All those results are reported in the supporting information for completeness.
Finally, at vanishing average density the kinetic energy contribution will become negligible with respect to the Coulomb repulsion.
Therefore, we expect that in this limit the positions at which the electrons localise will be equivalent to the positions for which classical point charges obtain their minimal energy.
We have numerically verified that this is indeed the case.
This comparison between the quantum and classical results can also be found in the supporting information.

In conclusion, we have proposed an accurate and efficient approach to study two-dimensional Wigner fragments from first principles.
It is based on the creation of a supercell that has the topology of a Clifford torus together with introduction of a renormalized distance in the Coulomb potential.
Our approach is fully quantum mechanical and makes no assumptions on the positions where the electrons will localise.
The density profiles we obtain, emerge naturally by minimising the total energy of the electrons confined to the fragments at low average density.
Our results indicate that, for large numbers of electrons, a hexagonal lattice structure will emerge.
We also conclude that, in general, not all the localised electrons are equivalent, even at very low density.
Finally, our work paves the way for studying other interesting properties of Wigner fragments, such as their magnetic properties, by including the spin degrees of freedom.
It could also be interesting to study three-dimensional Wigner fragments using the tools presented in this work.
\section{Supporting Information}
The density profiles at $r_s = 105$ of Wigner fragments  for $N$ electrons in a square Clifford supercell for all values of $N>1$ up to $N=20$ as well as the positions of the electrons obtained from classical calculations.
\section{Acknowledgments}
We thank the French ``Agence Nationale de la Recherche (ANR)'' for financial
support (Grant Agreements No. ANR-19-CE30-0011 and ANR-22-CE29-0001). This work
has been (partially) supported through the EUR grant NanoX n$^\circ$
ANR-17-EURE-0009 in the framework of the ``Programme des Investissements
d'Avenir''.

%\bibliography{Wigner.bib}

\providecommand{\latin}[1]{#1}
\makeatletter
\providecommand{\doi}
  {\begingroup\let\do\@makeother\dospecials
  \catcode`\{=1 \catcode`\}=2 \doi@aux}
\providecommand{\doi@aux}[1]{\endgroup\texttt{#1}}
\makeatother
\providecommand*\mcitethebibliography{\thebibliography}
\csname @ifundefined\endcsname{endmcitethebibliography}
  {\let\endmcitethebibliography\endthebibliography}{}

\end{document}